\documentclass{article}

\usepackage{amsmath}
\usepackage{amsthm, amssymb, amsfonts, amsbsy}
\usepackage{mathtools} 

\usepackage{bm}

\usepackage{graphicx}
\DeclareGraphicsExtensions{.pdf,.png}

\usepackage[margin=3cm]{geometry}
\usepackage{rotating}

\usepackage{lineno}
\usepackage{array}
\usepackage{authblk}
\usepackage{etoolbox}

\usepackage[draft=false, hyperindex=false, pagebackref=false, linkcolor=blue, citecolor=blue, colorlinks=true]{hyperref}

\usepackage[nocompress]{cite}
\usepackage[status=draft, mode=multiuser, margin]{fixme} 
\FXRegisterAuthor{ic}{ivc}{}
\FXRegisterAuthor{jt}{jxt}{}
\FXRegisterAuthor{rp}{rxp}{}
\fxusetheme{color}

\newcommand{\E}{E}
\newcommand{\Laplacian}{\Delta}

\begin{document}

\title{BeeNet: Reconstructing Flower Shapes from Electric Fields using Deep Learning}

\author[1, 2, 3]{Jake Turley} 
\author[4]{Ryan A Palmer} 
\author[2]{Isaac V Chenchiah}
\author[5]{Daniel Robert}

{\affil[1]{\small School of Biochemistry, University of Bristol, Biomedical Sciences Building, University Walk, Bristol BS8 1TD, UK}
 \affil[2]{\small School of Mathematics, University of Bristol, Fry Building, Woodland Road, Bristol, BS8 1UG, UK}
 \affil[3]{\small Mechanobiology Institute, National University of Singapore, 5A Engineering Drive 1, Singapore 117411}
 \affil[4]{\small School of Engineering Mathematics and Technology, University of Bristol, Ada Lovelace Building, University Walk, Bristol, BS8 1TW, UK}
 \affil[5]{\small School of Biological Sciences, University of Bristol, Life Sciences Building, 24 Tyndall Avenue, Bristol, BS8 1TQ, UK}}

\date{\today}

\maketitle

\begin{abstract}
Pollinating insects can obtain information from electric fields arising from flowers. The density and usefulness of electric information remain unknown. Here, we show that electric information can be used to reconstruct geometrical features of the field source. We develop an algorithm that infers the shapes of polarisable flowers from the electric field generated in response to a nearby charged arthropod. We computed the electric fields arising from arthropod–flower interactions for varying petal geometries, and used these data to train a deep learning U-Net model to recreate the floral shapes. The model accurately reconstructed diverse shapes, including more complex flower morphologies not included in training. Reconstruction performance peaked at an optimal arthropod–flower distance, indicating distance-dependent encoding of shape information. These findings indicate that electroreception can impart rich spatial detail, offering insights into the electric ecology of arthropods. Together, this work introduces a deep learning framework for solving the inverse electrostatic imaging problem, enabling object shape reconstruction directly from measured electric fields.
\end{abstract}

\section{Introduction}
Plants and arthropods have co-evolved over hundreds of millennia, giving rise to multiple sensory modalities, such as vision, hearing and smell, through which flora and fauna exchange valuable information. Of these many modalities, electric fields represent a subtle but ecologically important communication channel. The classic example is that of a positively charged bee causing a flower to electrically polarise upon approach. The magnitude, shape and variation of this so-called floral electric field are known to reflect the flower's geometry and resource abundance (e.g., pollen, nectar) \cite{clarke2013detection, woodburn2024electrostatic, Harris_2024}. Yet such ecological electric fields are easily perturbed by the act of measurement, hence their structure and informational content remain largely inaccessible to direct experimental observation. However, the physics of such electric fields is well defined and can be modelled mathematically~\cite{Griffiths_2023}. This creates an ideal setting in which machine learning may provide novel insights into a biological interaction that lies beyond the reach of conventional measurement.

Recent advances in machine learning offer powerful tools for inferring hidden structure in systems with indirect sensing~\cite{Chen_2020, Wang_2025}. Here, we hypothesise and test, by training models on physically modelled electric fields, that it is possible to recover information about the geometry of the objects that generate them. This approach illustrates how machine-learning models trained on physical simulations can extract latent structure from interactions that are not directly measurable, thereby offering a new way to investigate sensory ecology. Therefore, we ask and answer a fundamental question here: What information about a flower’s morphology and materials is contained within its electrical interaction with an approaching arthropod, and which features can be detected at a distance?

Machine-learning tools, particularly those from computer vision, are increasingly applied to biological systems. Deep learning models have been used to identify and segment cells~\cite{turley_deep_2024, Park_2023, BTRHGLE_2024, Singh_2020, Turley-Leong-Chan_2024}, and to classify dynamic behaviours such as division and extrusion~\cite{McDole_2018,turley_deep_2024-1,turley_deep_2024,villars_2023}. Recently, machine learning has been used to estimate the electric fields of complex objects, such as brains, producing models that are accurate and significantly faster than state-of-the-art physics-based simulators \cite{ahsan2025emulator, berger2025comprehensive, tanyel2024estimation}. In contrast to these forward models, we address the inverse problem: predicting the shape of an object from its electric field. Therefore, we adapt these computer vision methods to electroreception by introducing a deep learning model (``BeeNet'') with a 2D U-Net architecture, similar to standard segmentation networks but trained for a different objective. We demonstrate that this model can reconstruct flower shapes, solely from the electric fields generated by interactions between flowers and charged arthropods.

A substantial body of work has examined the electrical sensory abilities of arthropods and their behavioural consequences in communication~\cite{greggers2013reception}, dispersal~\cite{morley2018electric}, foraging~\cite{amador2017honey, khan2021electric},  parasitism~\cite{england2023static} and pollination~\cite{clarke2017bee}. Our investigation complements earlier work on the mechanics of electroreceptive hair sensing~\cite{palmer2023analysis, palmer2022mechanics,palmer2021analysis,palmer2023passive,Palmer-Chenchiah-Robert_2024, england2025electroreception}, including results showing that the location and shape of non-polarisable objects can be deduced from hair deflections~\cite{Palmer-Chenchiah-Robert_2024}. Flowers, however, are electrically polarisable \cite{clarke2013detection, woodburn2024electrostatic}, making their fields more complex to interpret and rendering theoretical analysis more challenging.

Such behavioural and mechanical evidence motivates the study of floral electrics as part of the broader ecology of plant–arthropod interactions. Although several studies have examined the electrostatics of plants~\cite{hunting2022synthetic,moyroud2017physics, montgomery2021bumblebee,  nanda2022study, molina2023electromagnetic, woodburn2024electrostatic}, many questions remain about the informational content of floral electric fields. In particular, it is unknown which aspects of a flower’s morphology and material properties are apparent in the field and therefore potentially detectable by electroreceptive organisms.

Here, we introduce a machine-learning framework that reconstructs flower geometry from mathematically modelled electric fields, allowing us to examine which structural features can be inferred from electric-field measurements and how robustly they influence the electric field. This approach provides a way to assess which morphological features are, in principle, present in floral electric fields, and suggests a potential strategy for inferring underlying physical interactions in systems where direct measurement is not possible. Beyond ecological electroreception, the same principles apply to non-contact sensing in robotics, long-range inverse problems in geophysics, and other settings where weak physical fields must be synthesised into structural information. Thus, our work, also demonstrates the potential capabilities of U-Nets in these and related areas.

\section{BeeNet}

\subsection{Modelling electric fields}

\paragraph{Floral electric fields.}
We consider a scenario in which a positively-charged arthropod, e.g., a bee, approaches an uncharged, polarisable flower ~\cite{clarke2013detection, clarke2017bee}. In response, an opposing electric field is generated in and around the flower. The structure and strength of this electric field, $E$, depend on the flower's shape and relative permittivity. It can be written as the gradient of the electric potential, $V$, 
\begin{subequations}
\begin{equation}
    E = -\nabla V.
\end{equation}

The flower is treated as a two-dimensional dielectric whose boundary, $\Gamma$, is described through an analytic function. Labelling the interior and exterior of the flower as $\Omega_1$ and $\Omega_2$, respectively, the electrical potential is defined as:
\begin{equation}
V = \begin{cases}
     V_1 & \text{ in } \Omega_1, \\
     V_2 & \text{ in } \Omega_2.
    \end{cases} 
\end{equation}
We thus obtain the following governing equations and boundary conditions, which couple the two domains through their common boundary:
\begin{alignat}{2}
\Laplacian V_1
 &= 0 && \text{ in } \Omega_1, \label{Meq:ilap} \\
\Laplacian V_2
 &= 0 && \text{ in } \Omega_2, \label{Meq:elap} \\
V_1
 &= V_2 && \text{ on } \Gamma, \label{Meq:dBC} \\
\frac{\partial V_2}{\partial n}
 &= \tilde{\epsilon} \frac{\partial V_1}{\partial n} &&  \text{ on } \Gamma. \label{Meq:nBC}
\end{alignat}
\end{subequations}
Here $\tilde{\epsilon}=\epsilon_1/\epsilon_2$ is the ratio of the absolute permittivities of the flower and the surrounding air, respectively. 
Boundary conditions~\eqref{Meq:dBC} and~\eqref{Meq:nBC} are a consequence of the electrical potential's continuity across the boundary and of Gauss' Law, respectively \cite{Griffiths_2023}. The far-field condition on the potential $V_2$ depends upon how $E$ is generated. 



\paragraph{Arthropod electric fields.}
Next, consider an arthropod, e.g., a bee, with a charge $\lambda$ which is several `petal radii', $L$, away from the flower centre. We assume that the arthropod is sufficiently small compared to the flower and, therefore, model it as a point charge located at $z_B$. Under these conditions, at a point $z$ somewhere in the domain, the electric potential associated with the arthropod in two-dimensional free space is given by
\begin{equation}\label{epotfree}
    V = -\frac{\lambda}{2\pi\epsilon}\log|z-z_B|,
\end{equation}
where the absolute permittivity $\epsilon$ is $\epsilon_1$ in $\Omega_1$ and $\epsilon_2$ in $z_2$. The potential $V$ is the quantity we compute in our datasets and to which our deep learning framework is applied. More details about the modelling approach can be found in~\cite{Harris_2024}.


\paragraph{Ecologically-relevant electrical interactions.}
From a sensory biology perspective, it is reasonable to assume that the arthropod does not detect itself and thus its contribution to the electric field should be subtracted. This is also computationally natural, since the arthropod's electric field dominates the flower's polarised field near the arthropod's location. What is ecologically relevant is the electrical interaction, i.e., how the flower perturbs the electric field generated by the charged arthropod. Hence, we compute the `perturbation field' of the flower, given by:
\begin{subequations}
\begin{align}
V_P &= V - V_A, \\
\E_P &= (E_{x}-E_{A,x}, E_{y}-E_{A,y}), \\
\| E_P \| &= \| E - E_A \|,
\end{align}
\end{subequations}
where the subscript $P$ indicates a perturbation value, $A$ (for ``arthropod'') indicates a scenario where only the charged arthropod is present, and the terms without subscripts are those of the modelled scenario, i.e., including both arthropod and flower.

Thus, the perturbation field may be interpreted as the `floral electrical information' from an arthropod's perspective, enabling us to assess and compare changes in field strength and structure as the flower distorts the arthropod's electric field.

\paragraph{Solving for the electric fields.}
The recently developed two-domain AAA-least squares (2D-AAA-LS) algorithm~\cite{Harris_2024} is used to solve the system of equations above. This method solves for the harmonic electric potential $V$ both inside and outside of a flower, subject to prescribed conditions at the flower-air interface and in the far-field. It is an extension of the AAA-LS method, which has become popular over the last few years having been applied to problems in mathematics, physics, chemistry and biology~\cite{nakatsukasa2023first,harris2023penguin,xue2023rational,kehry2023robust}, and has been extended~\cite{gopal2019solving,nakatsukasa2023first} to incorporate multiply connected domains~\cite{trefethen2020numerical,costa2023aaa} and a range of governing equations such as the Stokes~ \cite{brubeck2022lightning}, Poisson~\cite{harris2023penguin} and Helmholtz~\cite{gopal2019new} equations. Full details are given in the methods.

\subsection{Generating electric fields}

We are interested in how an uncharged dielectric flower, by polarising in the presence of an arthropod, alters the arthropod's electric field, thereby informing the arthropod about its shape and orientation. Therefore, we use the 2D-AAA-LS algorithm to generate perturbed electrical potentials and electric fields around a flower to investigate the possibility of reconstructing floral shapes from electric fields. We consider several scenarios and shapes that illustrate possible pollinator-plant interactions found in nature.

\begin{figure*}
\begin{center}
\includegraphics[width=\linewidth]{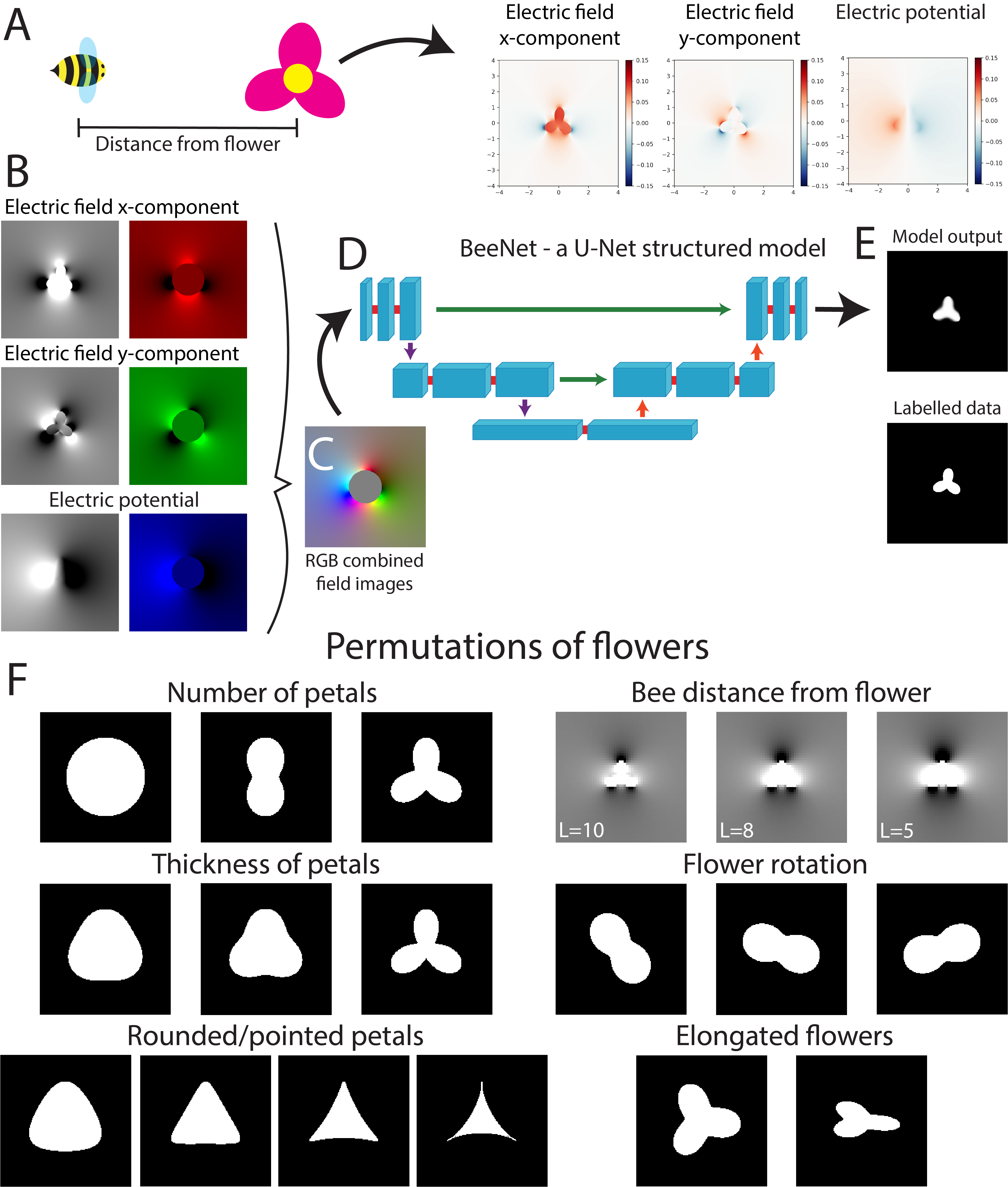} 
\caption{\small \sl A) Diagram of an arthropod and flower showing how their interaction alters the electric fields. B) The three resulting fields rendered as images: red and green represent perturbation fields in the x and y directions, and blue represents the electric potential. C) These three channels merged into a single RGB image. D) A simplified schematic of the BeeNet model based on standard U-Net architecture. E) Model predictions alongside the corresponding ground truth. F) The various flower configurations used during training.} \label{fig:diagram}
\end{center}
\end{figure*}

Overall, 1,979 datasets were produced, in which the floral shape, number of petals, flower orientation, floral permittivity and arthropod-flower distance were varied, as shown in Figure~\ref{fig:diagram}. In each, the solutions were evaluated on a square mesh to obtain the perturbation potential, $V_P$, and the perturbation electric field, $\E_P$. The field strength was normalised to lie in $[0,1]$. The electric field data is processed as a greyscale image, wherein negative values appear as dark regions and positive values as lighter regions (Figure~\ref{fig:diagram}B). After size reduction, the images are combined into an RGB image as in Figure~\ref{fig:diagram}C. These RGB images contain all the computed information about the flower's electric field in a format suitable for deep learning algorithms in computer vision. 

The maximum radius of each shape is $L$. Because the petal is clearly visible within this region (Figure~\ref{fig:diagram}B), we discard a disk of radius $1.1L$ from each image. This ensures that we retain only information about the surrounding electric fields produced by the arthropod-flower interaction, interpreted as the electrical sensory information propagated by the flower. Finally, each of these input images is matched with a binary petals mask, outlining the floral geometry (Figure~\ref{fig:diagram}C). Our goal is to train deep learning algorithms to reproduce these petal masks from the images containing information about the electric fields (Figure~\ref{fig:diagram}D).

\subsection{U-Net model for reconstructing flower shapes from electric fields}
To reconstruct flowers from their electric fields, we use a deep learning segmentation algorithm based on the U-Net architecture, a widely used and highly effective model~\cite{Ronneberger_2015}. Unlike traditional segmentation, which identifies objects with clearly defined visual boundaries, this more complex task involves inferring shape from electric field features at a distance. Nevertheless, we show that the same model architecture can be applied successfully. To this end, we converted an image classifier model ResNet101 into a U-Net structure via the fast.ai library's Dynamic U-Net class function~\cite{He_2016, howard_fastai_2018}. A simplified sketch of the model architecture is shown in Figure~\ref{fig:diagram}D, and the model output with its ground truth mask (labelled data) is shown in Figure~\ref{fig:diagram}E. 

Training deep learning models typically requires a large volume of data \cite{howard_fastai_2018, Hallou_2021}. Accordingly, we generated a dataset of 1,979 distinct floral variants, as described above and illustrated in Figure~\ref{fig:diagram}F. The model is trained solely on flowers with one to three rounded petals. An arthropod is considered the source of the electric field that electrically polarises the flower, with the arthropod-flower distance affecting the floral field and its strength. This parameter is varied from five to ten petal radii to generate test data. Lastly, we produced all of the above data for two values of relative floral permittivity, $\tilde{\epsilon}=10$ and $\tilde{\epsilon}=20$ (see~\eqref{Meq:nBC}), affecting the strength of the perturbed electric field. Consequently, with all these perturbations, an extensive training dataset was produced. 

To determine the model's effectiveness, we tested it on floral shapes and geometries not contained in the training data. This included four-petal flowers, to see if it could reconstruct these new shapes, and flowers with one to four petals with varied petal thickness and shape (e.g., pointed or rounded), rotation relative to the arthropod, petal length, and missing petals (Figure~\ref{fig:diagram}F).

\section{Results}

\subsection{BeeNet can infer flower structure from electric field.}

We initially evaluated the model using the validation data, which includeed flower shapes from the training data, but presented at a different, unseen orientation relative to the arthropod. Subsequently, we tested the model on a dataset with four petal types that it had not previously encountered.

\begin{figure} 
\begin{center}
\includegraphics[width=\linewidth]{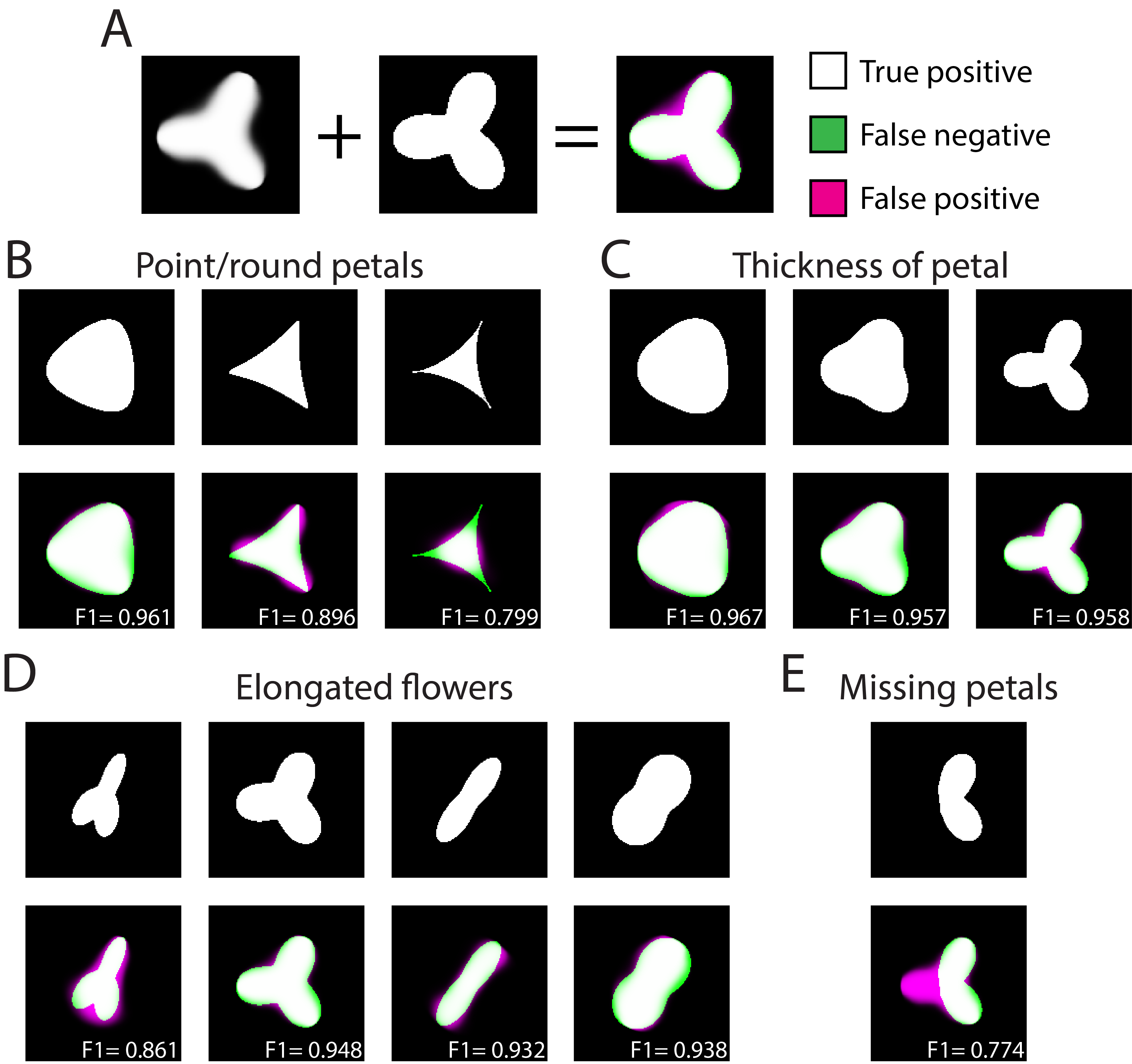} 
\caption{\small \sl A) The output of the model overlaying the ground truth. White pixels are correctly predicted (True positive). Green and magenta are the false negatives and false positives, respectively. B-E) The ground truth with the overlapping images displaying the accuracy for each of the shape perturbations} 
\label{fig:val}
\end{center}
\end{figure}

\subsubsection{The model's reconstruction of familiar flower shapes}

To demonstrate the model's effectiveness in reconstructing flowers, we overlay the output (magenta) and the ground-truth mask (green) as shown in Fig.~\ref{fig:diagram}A. The white regions are correctly identified areas of flower petals; these are the true positives, $T_p$, whilst green regions are false negatives, $F_n$, where the petals should be but were not captured by the model. The magenta regions are false positives, $F_p$, that the model incorrectly determines to be parts of the flowers. To quantify the effectiveness of the model, we will use the standard F1 score:
\begin{align}
    \text{F1} = \frac{2T_p}{2T_p + F_p + F_n} .
\end{align}
A perfect F1 score of 1 indicates that the image has been flawlessly reconstructed, whilst lower values of F1 indicate lower accuracy.
 
The validation dataset contained the same floral shapes as the training dataset, but rotated with respect to the arthropod, resulting in a configuration that had not been seen before. 

\begin{figure} 
\begin{center}
\includegraphics[width=\linewidth]{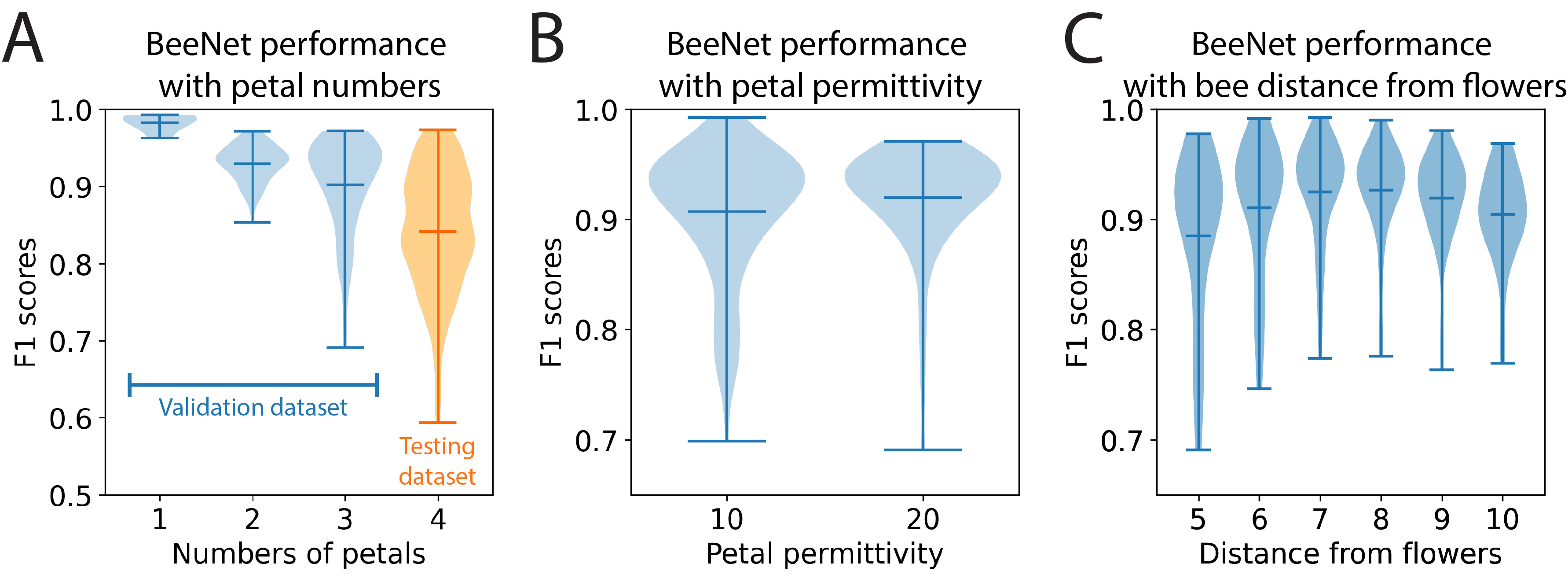} 
\caption{\small \sl A) Violin plot of the F1 score for each number of petals in flowers. The valuation dataset was used for flowers with 1-3 petals (blue) and test set was used for 4-petal flowers (orange). B) F1 score as a function of petal permeability, evaluated on the validation dataset. C) F1 score as a function of arthropod–flower distance, evaluated on the validation dataset.} 
\label{fig:quantify}
\end{center}
\end{figure}

On average, across all shapes in the validation set, we obtained an F1 score of 0.912 ($\pm$ 0.056 std). When comparing flowers with different petals, we found that the model reconstructed circular shapes most accurately, achieving an average F1 score of 0.983 ($\pm$ 0.01 std), Fig.~\ref{fig:quantify}A. For flowers with two and three petals, the scores dropped to 0.929 ($\pm$ 0.024 std) and 0.902 ($\pm$ 0.061 std), respectively. This indicates that as floral complexity increases, the model's ability to reconstruct geometry from its electric field at a distance decreases, though overall F1 scores remain high.

We examined whether altering the petals' permeability could modify the electric field and enhance the model's informational intake. 
In the validation dataset, raising the relative permeability of the petals from 10 to 20 led to a modest improvement in performance, increasing the average F1 score from 0.907 ($\pm$ 0.06 std) to 0.920 ($\pm$ 0.049 std), Fig.~\ref{fig:quantify}B. This small increase reflects an improved accuracy for flowers with the worst initial reconstructions. For example, for a relative permeability of 10, some floral shapes obtained scores ranging from 0.75 to 0.85, but for a relative permeability of 20, the performance of these flowers substantially improved (Fig. \ref{fig:quantify}B). This result is unsurprising, as a higher permittivity amplifies the electric field strength and therefore reflects the floral geometry more strongly at greater distances.

Since the flower polarises in the presence of the arthropod, the distance between the two influences the strength of the perturbation electric field and thus the information used by the model to reproduce the flower's structure. Based on this physical law, we expected that a closer arthropod, producing stronger floral electric fields, would lead to more distinct information for the reconstruction. 
After measuring the average F1 score from our database of petals, we found that there was a modest increase in the accuracy in the middle of our range, peaking at eight petal radii from the flower with a mean F1 score of $0.929$ ($\pm$ 0.044 std), shown in Fig.~\ref{fig:quantify}C. While at five and ten petal radii, mean values were $0.885$ ($\pm$ 0.076 std) and $0.904$ ($\pm$ 0.044 std), respectively. This result suggests that there are distances from the flower in which its electric field can be more readily used to determine its shape. 

We also investigated the role of petal shape to determine how changes in floral geometry affect reconstruction, e.g., pointed versus rounded petals or varying petal thicknesses. Flowers with rounded petals are closer to the circular geometry, which, as we have previously shown, are reconstructed with higher accuracy. Such petals had a high mean F1 score of $0.941$ ($\pm$ 0.026 std), whereas the model struggled to segment pointed petals with an average score of $0.785$ ($\pm$ 0.04 std). The largest inaccuracies occurred around the petal tips, which the model failed to recognise (Fig. \ref{fig:val}B). Thicker petals led to a small increase in accuracy (Fig.~\ref{fig:val}C), whilst petal elongation led to a small decrease in accuracy, mainly seen in three-petal flowers where there was a decrease in F1 score from $0.914$ ($\pm$ 0.055 std) to $0.876$ ($\pm$ 0.065 std) (Fig.~\ref{fig:val}D). Lastly, when we removed a petal from the three-petal flowers, the model incorrectly 're-added' it to the shape, resulting in a low F1 score (Fig.~\ref{fig:val}E). These observations are consistent with the electric fields and analysis presented in \cite{Harris_2024}, where pointed petals produced higher electric field values, which, however, diminished in strength over shorter distances than the rounded petals. This indicates a shorter propagation of geometric information in the floral electric fields for pointed petals, also confirmed here.

\subsubsection{BeeNet can reconstruct novel floral structures, albeit with reduced refinement.}

\begin{figure} 
\begin{center}
\includegraphics[width=\linewidth]{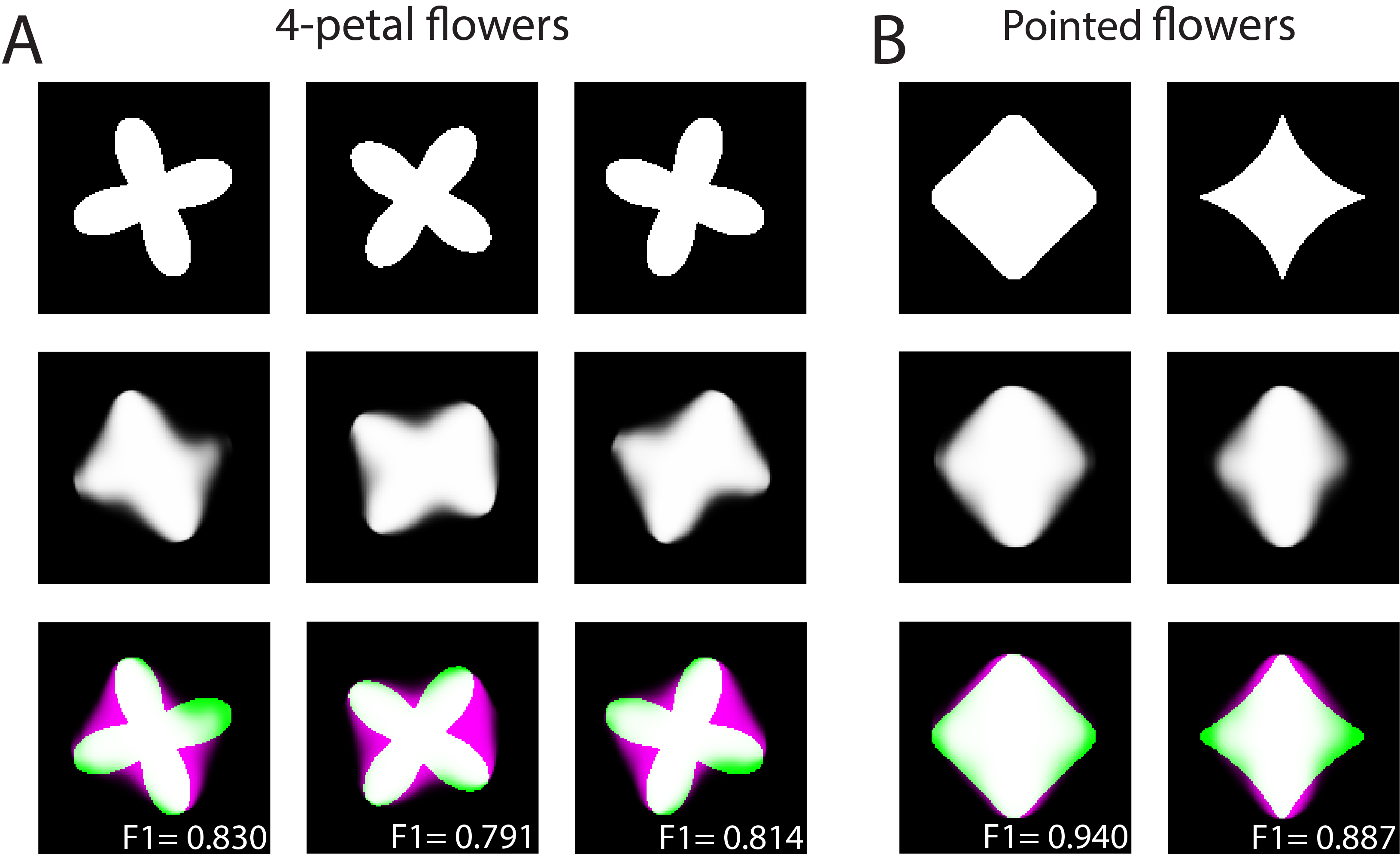} 
\caption{\small \sl A) Overlapping images displaying the accuracy for different rotations of the 4 petal flowers. B) 4 petal flowers with different levels of pointedness. (Green and magenta are the false negatives and false positives, respectively.)}
\label{fig:testing}
\end{center}
\end{figure}

We evaluated the model’s ability to reconstruct previously unseen flower shapes, testing its capacity to classify novel flower shapes rather than relying solely on learned geometries.
For this assessment, we selected four-petal flowers (Fig. \ref{fig:testing}).

Fig.~\ref{fig:testing}A shows that while the general four-petal structure is preserved, the reconstructions are less clear and sharp than those of previously seen flowers.
The average F1 score for four-petal flowers is 0.842 ($\pm$ 0.081 std), lower than the F1 score of 0.902 ($\pm$ 0.061 std) for three-petal flowers (Fig. \ref{fig:quantify}A), which aligns with the trend of decreasing accuracy for more complex and unfamiliar shapes.

Petals along the vertical axis are reconstructed more accurately than those along the horizontal axis, where some are missing or nearly lost. 
Given our modelled setting, where the arthropod is positioned to the left of each flower, we surmise that vertical components, which are perpendicular to the arthropod, provide more distinctive information than parallel components (Fig. \ref{fig:testing}B).
Despite these challenges, the model can segment various four-petal flower shapes, capturing their overall structure with a reasonable mean F1 score of 0.842 ($\pm$ 0.081 std).

\section{Discussion}

In this study, we have developed a deep learning model called BeeNet, that can reconstruct the shapes of flowers solely from their electric fields. BeeNet is shown to successfully segment shapes using the mid- to long-range strength and shape of floral electric fields with high accuracy, with an average F1 score of 0.911 ($\pm$ 0.049 std) in the validation set. 

In contrast to this focus on electrostatic floral signals, hitherto the majority of modelling has focused on the mechanics of hair sensors rather than the nature of the signals they receive. Indeed, this paper complements recent work that studied object reconstruction via  electromechanical actuation of sensory hairs \cite{Palmer-Chenchiah-Robert_2024}. Therein, the possibility that an observer could reconstruct the strength, structure and shape of a generating object was examined solely through the deflection of sensory hairs. However, the research question in this paper differs in two ways.
First, a `sensor agnostic' approach is followed. That is, whilst the field strength and potential are known, how this information is obtained is not considered. Therefore, our work investigates the problem of object reconstruction from an information-theoretic point of view, i.e., what in the nature of electric fields produces information that is potentially acquirable, interpretable and usable by an observer. Second, the information considered herein is that of the electric field about a flower due to polarisation caused by a static observer at a single location. In~\cite{Palmer-Chenchiah-Robert_2024}, however, an observer makes several observations of a statically charged object with no polarisation. Hence, the electrical information used to reconstruct the objects is different in both instances.

Returning to the role of floral geometry in electrostatic signals, crucially, BeeNet distinguishes between different thicknesses, shapes and numbers of the petals based solely on the electric field's variation and magnitude. This reveals the potential utility and nature of electrostatic sensory information within the wider sensory ecology of arthropods. Throughout the analysis, we show that distinct electrical `signatures' within the electric fields enable an electrosensitive observer to predict a flower's shape, orientation and distance. Our analyses also show how the shape, orientation and distance of the flower lead to strong qualitative and quantitative changes in the predictability of the electric field shape and strength. Thus, our results indicate that the broad variation of flower morphology \cite{byng2018phylogeny} can lead to distinct detectable and differentiable signals that convey perceptibly unique information about the flower~\cite{Harris_2024}.

We note that successful reconstruction indicates that the model has access to the relevant information in the electric field, but does not guarantee that the model has used this information in a physically meaningful way. Apparent success could arise from overfitting or from exploiting spurious correlations, so positive results should be interpreted as evidence of possible information content rather than definitive proof of its biological accessibility. To assess whether BeeNet generalises beyond the shapes seen during training, we evaluated it on four-petal flowers, despite training it only on one-, two-, and three-petal morphologies. The model recovered the fourth petal, albeit with lower spatial fidelity, indicating that it can infer certain structural features even outside the training distribution. 

Since this is a two-dimensional study, several aspects of flower morphology and heterogeneity, such as three-dimensionality and plant sex organs~\cite{byng2018phylogeny}, are not captured by our analysis. We anticipate that such features would further enhance the reconstruction of different floral shapes in three dimensions. In our study, we also found that there may be an arthropod-flower distance at which polarised floral fields more accurately convey information for reconstruction and identification (here, between six and nine radii). This relationship should be further studied, including whether there is also a dependence on the strength of the electric field in determining the optimal distance for flower identification. A dynamic equivalent of the model is also worth exploring. That is, to study the variation in the electric field as an arthropod moves around a flower and the possibility of identifying flowers from this variation.

As discussed in~\cite{palmer2023passive}, the strength, direction and influence of the electrical modality have distinct underlying physics compared to other senses.
Electrostatics, therefore, presents a particularly novel modality in the sensory world of arthropods since few other modalities, aside from vision, can elicit non-contact geometrical information at a distance for decision-making. 
Thus, electric fields present the possibility of a unique, information-rich and distinguishable information channel that is pervasive throughout the natural environment. 

Empirical and modelling studies are beginning to provide more evidence of the nuances of electric field sensing in arthropods and this modality's wider role in arthropod sensory ecology~\cite{clarke2013detection, sutton2016mechanosensory, morley2018electric, khan2021electric, england2021ecology, england2024prey}. Taking the results of this present study in the context of these experimental studies, other recent work on floral electrostatics~\cite{woodburn2024electrostatic, Harris_2024} and object reconstruction via sensory hairs~\cite{Palmer-Chenchiah-Robert_2024}, the potential power of electroreception and the information that can be conveyed and acquired is becoming clearer and more compelling. This spans the underlying physics that shapes natural electric fields, the electromechanical sensors that detect them, and the arthropod neural circuitry that interprets them--an inner world that machine learning in sensory biology is uniquely poised to illuminate.

\section{Methods}

\paragraph{Solving for the electric fields.}
Full details of the solution method are given in~\cite{Harris_2024}. In brief, solving the above equations yields the potential $V$ as the real part of an analytic function $F(z)$ that is approximated by rational functions, i.e., a polynomial plus the sum of singular terms from poles near the corners and cusps of the flower boundary. The function $F$ is then found using a linear leastsquares (LS) algorithm. 

Whilst other methods, such as finite element methods, can be used to solve the above problem, the 2D-AAA-LS algorithm is mesh-free, fast, and accurate: 
First, the 2D-AAA-LS algorithm provides an analytic solution based solely on the far-field conditions, the floral boundary and the interfacial conditions. Since the boundary is given by an analytic function, it can be quickly and efficiently changed via a simple adaptation of the generating parameters in an unsupervised manner. Mesh-based methods would require meshing for each geometry, requiring more time and less automation.
Second, the method is fast, typically solving each case in a fraction of a second. The speed of this method enables the efficient investigation of vast numbers of scenarios, making it well-suited to machine learning/AI approaches. Third, the result is highly accurate, producing relative errors in the order of $10^{-16}$ to $10^{-6}$ compared to analytical solutions and in contrast to relative errors of $10^{-2}$ in FEM approaches~\cite{Harris_2024}. Data generation was automated in MATLAB 2023b. 

We establish a maximum absolute value for the potential field $(V_{max})$ that trims only the highest values in the images (typically at the petal tips), while preserving high contrast across the majority of the field. This is needed because including extreme values in the normalisation process can obscure much of the signal in the remainder of the image. After trimming the highest absolute values, we apply the normalisation,
\begin{align}
\tilde{V_P} = 0.5 + \frac{V_P}{2V_{max}} \label{Meq:norm}
\end{align}
with similar normalisation for $E_{P,x}$ and $E_{P,y}$. In this manner, the potential and each component of the electric field were normalised to lie in $[0,1]$, which is a typical input range for images in deep learning algorithms.

The three field images are reduced in size from $801 \times 801$ to $401 \times 401$ pixels, maintaining a high-resolution of the electric fields while reducing the memory and computational power required for model training.

\paragraph{Datasets.}
The data were divided into three sets:

First, the training set comprised flowers with one to three petals, with all possible permutations depicted in Figure~\ref{fig:diagram}F. Every rotation of the flower relative to the arthropod was employed except one, which was reserved for the subsequent set.

Second, the validation set contains the same permutations as the training set, but at a single, distinct rotation relative to the arthropod. Essentially, it includes identical flower shapes as the training set, but in unfamiliar rotations.

Third, the test set consists exclusively of four-petal flowers. This will evaluate the model's ability to identify flowers, even though it has not been trained to recognize these specific shapes.

The data were augmented using the albumentations library \cite{Buslaev_2020}. The transforms used were Rotate, HorizontalFlip and VerticalFlip.

\paragraph{Network architecture and training models.}
We converted a ResNet101 model into a U-Net architecture via the Dynamic U-NET function from the fast.ai library \cite{He_2016, howard_fastai_2018, Ronneberger_2015, turley_deep_2024, turley_deep_2024-1}. 
The weights from the ResNet101 classifier were used to take advantage of transfer learning. BeeNet has 318,616,725 parameters and has 121 layers. This model has inputs of $401 \times 401 \times 3$. Source code is available at \url{https://github.com/turleyjm/BeeNet} and training data can be downloaded from \url{https://zenodo.org/records/16879470}. 

Paperspace’s gradient ML Platform was used to train the models. The machine used had either NVIDIA Quadra P5000 or P6000 GPU. We used an Adam optimization \cite{howard_fastai_2018}.

\medskip

\bibliography{References/ABbib, References/references_IC, References/references_JT}

@article{england2025electroreception,
author={England, Sam J and Palmer, Ryan A and O’Reilly, Liam J and Chenchiah, Isaac V and Robert, Daniel},
title={Electroreception in treehoppers: How extreme morphologies can increase electrical sensitivity},
journal={Proceedings of the National Academy of Sciences},
volume={122},
number={30},
pages={e2505253122},
year={2025},
month = jul,
doi={10.1073/pnas.2505253122}
}

@article{brubeck2022lightning,
author={Brubeck, P. D. and Trefethen, L. N.},
title={Lightning {S}tokes solver},
journal={SIAM Journal on Scientific Computing},
volume={44},
number={3},
pages={A1205-A1226},
year={2022},
doi={10.1137/21M1408579}
}

@inproceedings{costa2023aaa,
author = {Costa, S. and Trefethen, L. N.},
title = {{AAA}-least squares rational approximation and solution of {L}aplace problems},
editor = {Ademir Hujdurović and Klavdija Kutnar and Dragan Marušič and Štefko Miklavič and Tomaž Pisanski and Primož Šparl},
booktitle = {European Congress of Mathematics (Portorož, 20–26 June, 2021)},
year = {2023},
pages = {511-534},
doi = {10.4171/8ECM/16}
}

@article{gopal2019new,
author={Gopal, A. and Trefethen, L. N.},
title={New {L}aplace and {H}elmholtz solvers},
journal={Proceedings of the National Academy of Sciences},
volume={116},
number={21},
pages={10223-10225},
year={2019},
month=may,
doi={10.1073/pnas.1904139116}
}

@article{gopal2019solving,
author={Gopal, A. and Trefethen, L. N.},
title={Solving {L}aplace problems with corner singularities via rational functions},
journal={SIAM Journal on Numerical Analysis},
volume={57},
number={5},
pages={2074-2094},
year={2019},
doi={10.1137/19M125947X}
}

@article{harris2023penguin,
author={Harris, S. J. and McDonald, N. R.},
title={Penguin huddling: A continuum model},
journal={Acta Applicandae Mathematicae},
volume={185},
number={1},
pages={7},
year={2023},
month=jun
}

@article{kehry2023robust,
author={Kehry, M. and Klopper, W. and Holzer, C.},
title={Robust relativistic many-body {G}reen’s function based approaches for assessing core ionized and excited states},
journal={The Journal of Chemical Physics},
volume={159},
number={4},
pages={044116},
year={2023},
month=7,
doi={10.1063/5.0160265}
}

@misc{nakatsukasa2023first,
author={Nakatsukasa, Y. and Sete, O. and Trefethen, L. N.},
title={The first five years of the {AAA} algorithm},
note={arXiv:2312.03565},
year={2023}
}

@article{palmer2023passive,
author={Palmer, R. A. and Chenchiah, I. V. and Robert, D.},
title={Passive electrolocation in terrestrial arthropods: Theoretical modelling of location detection},
journal={Journal of Theoretical Biology},
volume={558},
pages={111357},
year={2023},
month=feb,
doi={10.1016/j.jtbi.2022.111357}
}

@article{palmer2023analysis,
author={Palmer, R. A. and O’Reilly, L. J. and Carpenter, J. and Chenchiah, I. V. and Robert, D.},
title={An analysis of time-varying dynamics in electrically sensitive arthropod hairs to understand real-world electrical sensing},
journal={Journal of the Royal Society Interface},
volume={20},
number={205},
pages={20230177},
year={2023},
month=aug,
doi={10.1098/rsif.2023.0177}
}

@article{trefethen2020numerical,
author={Trefethen, L. N.},
title={Numerical conformal mapping with rational functions},
journal={Computational Methods and Function Theory},
volume={20},
number={3},
pages={369-387},
year={2020},
month=jul
}

@inproceedings{xue2023rational,
author={Xue, C. and Xu, L and Wang, H and Liu, H and Yin, J and Li, X and Li, B},
title={A Rational Function Approximation Algorithm for the Frequency Sweep of Microwave Tube},
booktitle={24th International Vacuum Electronics Conference (IVEC)},
pages={1--2},
year={2023},
doi={10.1109/IVEC56627.2023.10157452}
}

@article{byng2018phylogeny,
author={Byng, J. W. and Smets, E F and van Vugt, R and Bidault, E and Davidson, C and Kenicer, G and Chase, M W and Christenhusz, M JM},
title={The phylogeny of angiosperms poster: A visual summary of {APG} {IV} family relationships and floral diversity},
journal={The Global Flora},
volume={4},
number={7},
pages={1--4},
year={2018}
}

@inproceedings{woodburn2024electrostatic,
author={Woodburn, FA and O’Reilly, LJ and Bentall, L and Robert, D},
title={Electrostatic detection and electric signalling in plants: Do flowers act as antennas?},
booktitle={Journal of Physics: Conference Series},
volume={2702},
pages={012012},
year={2024},
doi={10.1088/1742-6596/2702/1/012012}
}

@article{england2024prey,
  title={Prey can detect predators via electroreception in air},
  author={England, Sam J and Robert, Daniel},
  journal={Proceedings of the National Academy of Sciences},
  volume={121},
  number={23},
  pages={e2322674121},
  year={2024},
  publisher={National Academy of Sciences}
}

@article{hunting2022synthetic,
author={Hunting, E. R. and England, S. J. and Koh, K. and Lawson, D. A. and Brun, N. R and Robert, D},
title={Synthetic fertilizers alter floral biophysical cues and bumblebee foraging behavior},
journal={PNAS Nexus},
volume={1},
number={5},
pages={pgac230},
year={2022},
month=nov,
doi={10.1093/pnasnexus/pgac230}
}

@article{england2023static,
author={England, S. J. and Lihou, K. and Robert, D.},
title={Static electricity passively attracts ticks onto hosts},
journal={Current Biology},
volume = {33},
number = {14},
pages = {3041-3047},
year={2023},
doi={10.1016/j.cub.2023.06.021}
}

@article{sutton2016mechanosensory,
author={Sutton, G. P. and Clarke, D. and Morley, E. L. and Robert, D.},
title={Mechanosensory hairs in bumblebees (\textit{Bombus terrestris}) detect weak electric fields},
journal={Proceedings of the National Academy of Sciences},
volume={113},
number={26},
pages={7261-7265},
year={2016},
month=may,
doi={10.1073/pnas.1601624113}
}

@article{amador2017honey,
author={Amador, G. J. and Matherne, M. and Waller, D’A. and Mathews, M. and Gorb, S. N. and Hu, D. L.},
title={Honey bee hairs and pollenkitt are essential for pollen capture and removal},
journal={Bioinspiration and Biomimetics},
volume={12},
number={2},
pages={026015},
year={2017},
doi={10.1088/1748-3190/aa5c6e}
}

@article{morley2018electric,
author={Morley, E. L. and Robert, D.},
title={Electric fields elicit ballooning in spiders},
journal={Current Biology},
volume={28},
number={14},
pages={2324-2330.e2},
year={2018},
month=jul,
doi={10.1016/j.cub.2018.05.057}
}

@article{clarke2017bee,
author={Clarke, D. and Morley, E. and Robert, D.},
title={The bee, the flower, and the electric field: Electric ecology and aerial electroreception},
journal={Journal of Comparative Physiology A},
volume={203},
number={9},
pages={737--748},
year={2017},
month=jun,
doi={10.1007/s00359-017-1176-6}
}

@article{clarke2013detection,
author={Clarke, D. and Whitney, H. and Sutton, G. and Robert, D.},
title={Detection and learning of floral electric fields by bumblebees},
journal={Science},
volume={340},
number={6128},
pages={66-69},
year={2013},
month=feb,
doi={10.1126/science.1230883}
}

@article{palmer2021analysis,
author={Palmer, R. A. and Chenchiah, I. V. and Robert, D.},
title={Analysis of aerodynamic and electrostatic sensing in mechanoreceptor arthropod hairs},
journal={Journal of Theoretical Biology},
volume={530},
pages={110871},
year={2021},
month=dec,
doi={10.1016/j.jtbi.2021.110871}
}

@article{greggers2013reception,
author={Greggers, U. and Koch, G. and Schmidt, V. and D{\"u}rr, A. and Floriou-Servou, A. and Piepenbrock, D. and G{\"o}pfert, M. C. and Menzel, R.},
title={Reception and learning of electric fields in bees},
journal={Proceedings of the Royal Society B: Biological Sciences},
volume={280},
number={1759},
pages={20130528},
year={2013},
month=may,
doi = {10.1098/rspb.2013.0528}
}

@article{england2021ecology,
author={England, S. J. and Robert, D.},
title={The ecology of electricity and electroreception},
journal={Biological Reviews},
volume={97},
number = {1},
pages={383-413},
year={2021},
month=feb,
doi={10.1111/brv.12804}
}

@article{palmer2022mechanics,
author={Palmer, R. A. and Chenchiah, I. V. and Robert, D.},
title={The mechanics and interactions of electrically sensitive mechanoreceptive hair arrays of arthropods},
journal={Journal of the Royal Society Interface},
volume={19},
number={188},
pages={20220053},
year={2022},
doi={10.1098/rsif.2022.0053}
}

@article{khan2021electric,
author={Khan, S. A. and Khan, K. A. and Kubik, S. and Ahmad, S. and Ghramh, H. A. and Ahmad, A. and Skalicky, M. and Naveed, Z. and Malik, S. and Khalofah, A. and others},
title={Electric field detection as floral cue in hoverfly pollination},
journal={Scientific Reports},
volume={11},
pages={18781},
year={2021},
month=sep,
doi={10.1038/s41598-021-98371-4}
}

@article{molina2023electromagnetic,
author={Molina-Montenegro, M. A. and Acu{\~n}a-Rodr{\'\i}guez, I S and Ballesteros, G I and Baldelomar, M and Torres-D{\'\i}az, C and Broitman, B R and V{\'a}zquez, D P},
title={Electromagnetic fields disrupt the pollination service by honeybees},
journal={Science Advances},
volume={9},
number={19},
pages={eadh1455},
year={2023},
month=may,
doi={10.1126/sciadv.adh1455}
}

@article{montgomery2021bumblebee,
author={Montgomery, C. and Vuts, J. and Woodcock, C. M. and Withall, D. M. and Birkett, M. A. and Pickett, J. A. and Robert, D.},
title={Bumblebee electric charge stimulates floral volatile emissions in Petunia integrifolia but not in Antirrhinum majus},
journal={The Science of Nature - Naturwissenschaften},
volume={108},
pages={44},
year={2021},
month=sep
}

@article{moyroud2017physics,
author={Moyroud, E. and Glover, B. J.},
title={The physics of pollinator attraction},
journal={New Phytologist},
volume={216},
number={2},
pages={350-354},
year={2017},
month=oct,
doi={10.1111/nph.14312}
}

@article{nanda2022study,
author={Nanda, I. and De, R. and others},
title={Study of Electromagnetic Radiation on Flower},
journal={Matrix Science Mathematic},
volume={6},
number={2},
pages={58-63},
year={2022},
doi={10.26480/msmk.02.2022.58.63}
}

@book{Griffiths_2023,
author={Griffiths, David J.}, 
title={Introduction to Electrodynamics}, 
publisher={Cambridge University Press}, 
place={Cambridge},
edition={5},
year={2023}
}

@article{McDole_2018,
author = {McDole, Katie and Guignard, Léo and Amat, Fernando and Berger, Andrew and Malandain, Grégoire and Royer, Loïc A. and Turaga, Srinivas C. and Branson, Kristin and Keller, Philipp J.},
title = {\textit{In Toto}  and Reconstruction of Post-Implantation Mouse Development at the Single-Cell Level},
journal = {Cell},
volume = {175},
number = {3},
year = {2018},
month = oct,
pages = {859-876.e33},
doi = {10.1016/j.cell.2018.09.031}
}

@inproceedings{He_2016,
author={He, Kaiming and Zhang, Xiangyu and Ren, Shaoqing and Sun, Jian},
title={Deep Residual Learning for Image Recognition}, 
booktitle={2016 IEEE Conference on Computer Vision and Pattern Recognition (CVPR)}, 
year={2016},
pages={770-778},
doi={10.1109/CVPR.2016.90}
}

@article{Hallou_2021,
author = {Hallou, Adrien and Yevick, Hannah G. and Dumitrascu, Bianca and Uhlmann, Virginie},
title = {Deep learning for bioimage analysis in developmental biology},
journal = {Development},
volume = {148},
number = {18},
pages = {dev199616},
year = {2021},
month = sep,
doi = {10.1242/DEV.199616}
}

@article{Buslaev_2020,
author = {Buslaev, Alexander and Iglovikov, Vladimir I. and Khvedchenya, Eugene and Parinov, Alex and Druzhinin, Mikhail and Kalinin, Alexandr A.},
title = {Albumentations: Fast and Flexible Image Augmentations},
journal = {Information},
volume = {11},
number = {2},
article-number = {125},
year = {2020},
doi = {10.3390/info11020125}
}

@article{Park_2023,
author = {Park, Juyeon and Bai, Bijie and Ryu, DongHun and Liu, Tairan and Lee, Chungha and Luo, Yi and Lee, Mahn Jae and Huang, Luzhe and Shin, Jeongwon and Zhang, Yijie and Ryu, Dongmin and Li, Yuzhu and Kim, Geon and Min, Hyun-seok and Ozcan, Aydogan and Park, YongKeun},
title = {Artificial intelligence-enabled quantitative phase imaging methods for life sciences},
journal = {Nature Methods},
volume = {20},
year = {2023},
month = oct,
pages = {1645--1660},
doi = {10.1038/s41592-023-02041-4}
}

@InProceedings{Ronneberger_2015,
author = "Ronneberger, Olaf and Fischer, Philipp and Brox, Thomas",
title = "U-Net: Convolutional Networks for Biomedical Image Segmentation",
editor = "Navab, Nassir and Hornegger, Joachim and Wells, William M. and Frangi, Alejandro F.",
booktitle = "Medical Image Computing and Computer-Assisted Intervention (MICCAI 2015)",
pages="234-241",
year = "2015"
}

@article{Singh_2020,
author = {Singh, Satya P. and Wang, Lipo and Gupta, Sukrit and Goli, Haveesh and Padmanabhan, Parasuraman and Gulyás, Balázs},
title = {{3D} Deep Learning on Medical Images: A Review},
journal = {Sensors},
volume = {20},
number = {18},
year = {2020},
month = sep,
pages = {5097},
doi = {10.3390/s20185097}
}

@article{Turley-Leong-Chan_2024,
author = {Turley, Jake and Leong, Kim Whye and Chan, Chii Jou},
title = {Novel imaging and biophysical approaches to study tissue hydraulics in mammalian folliculogenesis},
journal = {Biophysical Reviews},
volume = {16},
pages = {625-637},
year = {2024},
month = oct,
doi = {10.1007/s12551-024-01231-4}
}

@article{Villars_2023,
author = {Villars, Alexis and Letort, Gaëlle and Valon, Léo and Levayer, Romain},
title = {{DeXtrusion}: Automatic recognition of epithelial cell extrusion through machine learning in vivo},
journal = {Development},
volume = {150},
number = {13},
pages = {dev201747},
year = {2023},
month = jul,
doi = {10.1242/dev.201747}
}

@misc{Wang_2025,
author = {Wang, Ji-Yuan and Lou, Xin-Yue and Zhang, Liang and Wang, Yun-Chuan and Pan, Xiao-Min},
title = {A Deep Learning Scheme of Electromagnetic Scattering from Scatterers with Incomplete Profiles},
year = {2025},
month = may,
doi = {10.48550/arXiv.2505.02086},
note = {arXiv:2505.02086 [cs]}
}

@article{Chen_2020,
author = {Chen, Xudong and Wei, Zhun and Li, Maokun and Rocca, Paolo},
title = {A Review of Deep Learning Approaches for Inverse Scattering Problems},
journal = {Progress in Electromagnetics Research},
volume = {167},
pages = {67-81},
year = {2020},
doi = {10.2528/PIER20030705}
}

@article{BTRHGLE_2024,
    author = {Barone, Vanessa and Tagua, Antonio and Rom\'an, Jesus \'A. Andr\'es-San and Hamdoun, Amro and Garrido-Garc\'ia, Juan and Lyons, Deirdre C. and Escudero, Luis M.},
    title = "{Local and global changes in cell density induce reorganisation of 3D packing in a proliferating epithelium}",
    journal = {Development},
    volume = {151},
    number = {20},
    pages = {dev202362},
    year = {2024},
    month = {05},
    doi = {10.1242/dev.202362},
    url = {https://doi.org/10.1242/dev.202362}
}

@article{Palmer-Chenchiah-Robert_2024,
author = {Ryan A Palmer and Isaac V Chenchiah and Daniel Robert},
title = {Sensing electrical environments: mechanical object reconstruction via electrosensors},
journal = {Journal of Physics A: Mathematical and Theoretical},
volume = {57},
number = {38},
pages = {385601},
year = {2024},
month = sep,
doi = {10.1088/1751-8121/ad6f80}
}

@article{Harris_2024,
author = {Harris, Samuel J. and Palmer, Ryan A. and McDonald, N. R.},
title = {Modeling Floral and Arthropod Electrostatics Using a Two-Domain {AAA-Least} Squares Algorithm},
journal = {SIAM Journal on Applied Mathematics},
volume = {85},
number = {2},
pages = {916-944},
year = {2025},
doi = {10.1137/24M1638021}
}

@article{ahsan2025emulator,
  title={EMulator: Rapid Estimation of Complex-valued Electric Fields using a U-Net Architecture},
  author={Ahsan, Fatima and Luzi, Lorenzo and Barainuk, Richard G and Sheth, Sameer A and Goodman, Wayne and Aazhang, Behnaam},
  journal={arXiv preprint arXiv:2505.02095},
  year={2025}
}

@article{berger2025comprehensive,
  title={Comprehensive evaluation of U-Net based transcranial magnetic stimulation electric field estimations},
  author={Berger, Taylor A and Mantell, Kathleen and Haigh, Zachary and Perera, Nipun and Alekseichuk, Ivan and Opitz, Alexander},
  journal={Scientific reports},
  volume={15},
  number={1},
  pages={12204},
  year={2025},
  publisher={Nature Publishing Group UK London}
}

@inproceedings{tanyel2024estimation,
  title={Estimation of Two Dimensional Electric Field Distribution Through Deep Learning: Preliminary Study},
  author={Tanyel, Toygar and Yildiz, Gulsah and Aydinalp, Cemanur and {\"O}ks{\"u}z, {\.I}lkay},
  booktitle={2024 32nd Signal Processing and Communications Applications Conference (SIU)},
  pages={1--4},
  year={2024},
  organization={IEEE}
}

@article{turley_deep_2024,
	title = {Deep learning reveals a damage signalling hierarchy that coordinates different cell behaviours driving wound re-epithelialisation},
	volume = {151},
	copyright = {http://creativecommons.org/licenses/by/4.0},
	issn = {0950-1991, 1477-9129},
	url = {https://journals.biologists.com/dev/article/151/18/dev202943/362123/Deep-learning-reveals-a-damage-signalling},
	doi = {10.1242/dev.202943},
	language = {en},
	number = {18},
	urldate = {2024-10-25},
	journal = {Development},
	author = {Turley, Jake and Robertson, Francesca and Chenchiah, Isaac V. and Liverpool, Tanniemola B. and Weavers, Helen and Martin, Paul},
	month = sep,
	year = {2024},
	pages = {dev202943},
}

@article{turley_deep_2024-1,
	title = {Deep learning for rapid analysis of cell divisions in vivo during epithelial morphogenesis and repair},
	volume = {12},
	copyright = {All rights reserved},
	issn = {2050-084X},
	url = {https://elifesciences.org/articles/87949},
	doi = {10.7554/eLife.87949},
	language = {en},
	urldate = {2024-10-25},
	journal = {eLife},
	author = {Turley, Jake and Chenchiah, Isaac V and Martin, Paul and Liverpool, Tanniemola B and Weavers, Helen},
	month = sep,
	year = {2024},
	pages = {RP87949},
}

@misc{howard_fastai_2018,
	title = {Fast.ai},
	author = {Howard, J. and {others}},
	year = {2018},
}

\end{document}